\documentclass[aps,prd,twocolumn,groupedaddress,showpacs]{revtex4}

\usepackage{graphicx}% Include figure files

\usepackage{dcolumn}% Align table columns on decimal point

\usepackage{bm}% bold math

\renewcommand\({\left(}
\renewcommand\){\right)}
\renewcommand\[{\left[}
\renewcommand\]{\right]}

\newcommand{\ra}{\rightarrow}

\def\lsim{\raise 0.4ex\hbox{$<$}\kern -0.8em\lower 0.62
ex\hbox{$\sim$}}

\def\gsim{\raise 0.4ex\hbox{$>$}\kern -0.7em\lower 0.62
ex\hbox{$\sim$}}

\def\lbar{{\hbox{$\lambda$}\kern -0.7em\raise 0.6ex
\hbox{$-$}}}

\newcommand\eq[1]{eq.~(\ref{#1})}

\newcommand\p{\partial}

\newcommand\ee{\end{equation}}
\newcommand\be{\begin{equation}}
\def\bea{\begin{array}}
\def\eea{\end{array}}\def\ea{\end{array}}
\newcommand\ees{\end{eqnarray}}
\newcommand\bees{\begin{eqnarray}}
\def\nn{\nonumber}

\def\v#1{\hbox{\boldmath$#1$}}

\def\vchi{\v{\chi}}

\def\D{\Delta}

\def\b{\beta}

\def\s{\sigma}
\def\g{\gamma}

\def\d{\delta}

\def\dslash{\hspace{-1mm}\not{\hbox{\kern-2pt $\partial$}}}
\def\Dslash{\not{\hbox{\kern-4pt $D$}}}
\def\pslash{\not{\hbox{\kern-2.1pt $p$}}}
\def\kslash{\not{\hbox{\kern-2.3pt $k$}}}
\def\qslash{\not{\hbox{\kern-2.3pt $q$}}}

\newcommand{\vx}{{\bf x}}

\newcommand{\bdot}{{\bf\cdot}}

\def\p1{{\bf p}_1}
\def\p2{{\bf p}_2}
\def\k1{{\bf k}_1}
\def\k2{{\bf k}_2}

\newcommand{\emn}{\eta_{\mu\nu}}

\newcommand{\gmn}{g_{\mu\nu}}

\newcommand{\hmn}{h_{\mu\nu}}

\newcommand{\bhmn}{\bar{h}_{\mu\nu}}

\newcommand{\hatr}{\hat{\bf r}}

\newcommand{\dddM}{\kern 0.2em \raise 1.9ex\hbox{$...$}\kern -1.0em \hbox{$M$}}
\newcommand{\dddQ}{\kern 0.2em \raise 1.9ex\hbox{$...$}\kern -1.0em \hbox{$Q$}}
\newcommand{\dddI}{\kern 0.2em \raise 1.9ex\hbox{$...$}\kern -1.0em\hbox{$I$}}
\newcommand{\dddJ}{\kern 0.2em \raise 1.9ex\hbox{$...$}\kern-1.0em
\hbox{$J$}}
\newcommand{\dddcalJ}{\kern 0.2em \raise 1.9ex\hbox{$...$}\kern-1.0em
\hbox{${\cal J}$}}

\newcommand{\dddO}{\kern 0.2em \raise 1.9ex\hbox{$...$}\kern -1.0em
\hbox{${\cal O}$}}
\def\dddz{\raise 1.5ex\hbox{$...$}\kern -0.8em \hbox{$z$}}
\def\dddd{\raise 1.8ex\hbox{$...$}\kern -0.8em \hbox{$d$}}
\def\dddbd{\raise 1.8ex\hbox{$...$}\kern -0.8em \hbox{${\bf d}$}}
\def\ddbd{\raise 1.8ex\hbox{$..$}\kern -0.8em \hbox{${\bf d}$}}
\def\dddx{\raise 1.6ex\hbox{$...$}\kern -0.8em \hbox{$x$}}

\newcommand{\msun}{M_{\odot}}

\newcommand{\mpl}{M_{\rm Pl}}

\newcommand{\bppn}{\beta_{\rm PPN}}
\newcommand{\gppn}{\gamma_{\rm PPN}}

\begin{document}

\title{Extracting  the three- and four-graviton vertices
from binary pulsars \\and coalescing binaries}

\author{Umberto Cannella$^{(1)}$, Stefano Foffa$^{(1)}$, 
Michele Maggiore$^{(1)}$, Hillary Sanctuary$^{(1)}$ and 
Riccardo Sturani$^{(2,3)}$}

\affiliation{$(1)$ D\'epartement de Physique Th\'eorique, Universit\'e de 
             Gen\`eve, CH-1211 Geneva, Switzerland\\
$(2)$ Istituto di Fisica, Universit\`a di Urbino, I-61029 Urbino, Italy\\
$(3)$ INFN, Sezione di Firenze, I-50019 Sesto Fiorentino, Italy}

%\email{e-mail:Umberto.Cannella@unige.ch, Stefano.Foffa@unige.ch, 
%Michele.Maggiore@unige.ch, Hillary.Sanctuary@unige.ch, 
%Riccardo.Sturani@uniurb.it}

%\date{\today}

\begin{abstract}
Using a formulation of the post-Newtonian  expansion in terms of Feynman 
graphs, we discuss how various tests of General Relativity (GR) can be 
translated into measurement of the three- and four-graviton vertices.
In problems involving only the conservative dynamics of a system, a
deviation of the three-graviton vertex from the GR prediction is
equivalent, to lowest order, to the introduction of the parameter $\bppn$
in the parametrized post-Newtonian formalism, and its strongest bound comes
from lunar laser ranging, which measures it at the 0.02\% level. 
Deviation of the three-graviton vertex from the GR
prediction, however, also affects the radiative sector of the theory. We
show that  the timing of the  Hulse-Taylor binary
pulsar provides a bound on the deviation of the three-graviton vertex 
from the GR prediction at the 0.1\% level.
For coalescing binaries at  interferometers we find that,  
because of degeneracies with other parameters in the template 
such as mass and spin, the effects of modified
three- and four-graviton vertices is just to induce an error in the
determination of these parameters and, at least in the restricted PN
approximation,  it is not possible to use coalescing binaries for
constraining  deviations of the vertices
from the GR prediction.

\end{abstract}



%\keywords{}


\maketitle

\section{Introduction}

Binary pulsars,  such as the Hulse-Taylor \cite{HulseTaylor} and 
the  double pulsar  \cite{Bug03}, are
wonderful laboratories for testing  General
Relativity (GR).  They have given the first experimental
confirmation of the existence of gravitational radiation
\cite{TW82,WeisT04},  provide stringent tests of GR
and allow
for comparison with alternative theories of gravity, such as scalar-tensor
theories~\cite{DT91,DamT2,TWDW,Kramer:2006nb,DF92,DF93,Will94,DF96,DF96b,DF98}
(see \cite{Stairs,Will:2008li,Maggiore:1900zz} for reviews). 
Another very sensitive probe of the non-linearities of GR is given
by the  gravitational wave (GW) emission during the last stages of the
coalescence of compact binary systems made of black holes
and/or neutron stars, which is one of the
most promising signals for GW interferometers such as LIGO and Virgo,
especially in their advanced stage, and for the space interferometer LISA.
Various investigations have been devoted to the possibility of using
the observation of coalescing binaries at GW interferometers to probe 
non-linear aspects of
GR \cite{Will94,DF98,BS1,BS2,Will98,EBW1,EBW2,Arun:2006yw,Arun:2006hn}.

Compact binary systems  probe both  the radiative sector of the theory, through
the emission of gravitational radiation, and  the
non-linearities intrinsic to GR which are already present  in the
conservative part of the Lagrangian.
In a field-theoretical language, these non-linearities can be traced
to the non-abelian vertices of the theory, such as the three-
and four-graviton vertices. It is therefore natural to ask
whether from binary pulsars or from future observations of coalescing
binaries at interferometers one can extract a measurement of these
vertices, much in the same spirit in which the triple
and quartic gauge boson couplings have been measured at LEP2 and at the 
Tevatron~\cite{DeRujula:1978bz,Ellison:1998uy,Dzero,Schael:2004tq}.

In this paper we tackle this question.  The organization of the paper
is as follows. In Section~\ref{sect:tag} we discuss how to ``tag''
the contribution of the three- and four-graviton vertices to various
observables in a consistent and gauge-invariant
manner, and we compare it with other approaches, such as the
parametrized post-Newtonian (PPN) formalism~\cite{Will:2008li}.
In particular, we
find that the introduction of a modified three-graviton vertex
corresponds -- in the conservative sector of the theory and at first 
Post-Newtonian order (1PN) -- to the
introduction of a value for the PPN parameter $\bppn$ different from the
value $\bppn=1$ of GR. However, a modified three-graviton vertex also
affects the radiative sector of the theory, which is not the case for
the PPN parameter $\bppn$. We also discuss  subtle issues related to
the possible breaking of gauge invariance which takes place when one modifies
the vertices of the theory.
In Section~\ref{sect:comp} we present our computations with
modified vertices, and in Section~\ref{sect:exp} we compare these computations
with experimental results obtained from the timing of
binary pulsars and 
with what can be expected from the detection of gravitational waves
(GWs) at ground-based interferometers or with
the space interferometer LISA. Section~\ref{sect:concl} contains our
conclusions.

\section{Tagging the three- and four-graviton vertices}\label{sect:tag}

Our aim is to quantify how well the
non-linearities of GR can be tested by various existing or planned 
experiments/observations. Historically, there have been several approaches
to this problem and, 
basically, one can identify two complementary strategies. The
first is to develop a purely phenomenological approach in which
deviations from GR are expressed in terms of a number of parameters,
without inquiring at first whether such a
deformation of GR can emerge from a fundamental theory.
An example of such an approach is the
parametrized post-Newtonian (PPN) formalism. In its simpler version,
it consists of writing the 1PN metric generated by a source, treated
as a perfect fluid with density $\rho(\vx)$ and velocity field
$v(\vx)$, 
in the form
\bees
g_{00}&=&-1+2U-2\bppn U^2\, ,\label{g00}\\
g_{0i}&=&-\frac{1}{2}(4\gppn+3)V_i\, ,\\
g_{ij}&=&(1+2\gppn U)\d_{ij}\, ,\label{gij}
\ees
where
\bees
U(\vx )&=&\int d^3x'\, \frac{\rho(x')}{|\vx-\vx'|}\, ,\\
V_i(\vx )&=&\int d^3x'\, \frac{\rho(x')v_i(x')}{|\vx-\vx'|}\, ,
\ees
and the standard PPN gauge has been used~\cite{Willbook,Will:2008li}
(we use units $c=1$).
General Relativity corresponds to $\bppn=1$ and $\gppn =1$. More
phenomenological parameters can be introduced  by working at higher PN
orders, see \cite{Will:2008li}.
One then investigates how deviations of $\bppn$ and $\gppn$ from
their GR values affect various experiments.
Writing $\bppn=1+\bar{\b}$ and $\gppn=1+\bar{\g}$, the best current
limits (at 68\% c.l.) are
\be
\bar{\g}=(2.1\pm 2.3)\times 10^{-5}\, 
\ee
from
the Doppler tracking of the Cassini spacecraft, and
\be\label{Cassini44}
4\bar{\beta}-\bar{\g}=(4.4\pm 4.5)\times 10^{-4}\, 
\ee
from lunar laser ranging. 
This bound comes from the
Nordtvedt effect, i.e. from the fact that, in a theory with $\bar{\beta}$ 
and $\bar{\g}$ generic,  the weak equivalence principle is violated and the Earth and the Moon can fall toward the Sun with different accelerations, which depend  on their 
gravitational self-energy. The effect is studied by monitoring the 
Earth-Moon distance with lunar laser ranging.
The perihelion shift of Mercury gives instead the bound
$|\bar{\b}|<3\times 10^{-3}$~\cite{Will:2008li,pdg}.

To get these bounds, we do not need to know the fundamental
theory that gives rise to values of $\bppn$ and $\gppn$ 
that differ from their
GR values. However, it is of course interesting to see that consistent  
field theories exist
that give rise to values of $\bppn$ and $\gppn$
different from one. For instance,
a Brans-Dicke theory with parameter $\omega_{\rm BD}$ gives
$\bppn=1$ and $\gppn =(1+\omega_{\rm BD})/(2+\omega_{\rm BD})$, with the GR
value $\gppn=1$ recovered for $\omega_{\rm BD}\ra\infty$, while  more
general tensor-scalar theories can produce both $\gppn\neq 1$ and 
$\bppn\neq 1$. However, in the PPN approach, one can also explore 
other possibilities,
such as
PPN parameters that correspond to preferred-frame effects or
to violation of the conservation of total momentum, which are not
necessarily well-motivated in terms of  current field-theoretical
ideas on possible extensions or UV completions of GR.
It is also important to observe 
that the parameters $\bppn$ and $\gppn$ are gauge-invariant,
and therefore observables, because they have been defined with respect
to a specific gauge, namely the standard PPN 
gauge in which the metric takes the 
form~(\ref{g00})--(\ref{gij}).

A second, complementary, approach to the problem is to study a specific
class of field-theoretical extensions of GR. A typical well-motivated
example 
is provided by multiscalar-tensor theories. 
These have been studied
in detail and compared with experimental tests of relativistic gravity
in refs.~\cite{DamT2,DF92,DF93,Will94,DF96,DF96b,DF98}. This approach has
the advantage that one is testing a specific and well-defined
fundamental theory. On the other hand, an experimental  bound on the
parameters of a given scalar-tensor theory, such as for instance the
bound $\omega_{\rm BD}>40000$ on the parameter $\omega_{\rm BD}$ of Brans-Dicke
theory obtained from the tracking of the Cassini 
spacecraft~\cite{Bertotti:2003rm} is,  strictly speaking, only a
statement about that particular extension of GR and not about GR itself.

In this paper we quantify how well GR performs with respect
to experiments of relativistic gravity by studying how much these
experiments constrain the values of the non-abelian vertices of the
theory, in particular the three-graviton vertex and the four-graviton
vertex. We proceed as follows. After choosing a gauge
(the De~Donder gauge, corresponding to harmonic coordinates) we multiply the 
three-graviton vertex by a factor $(1+\b_3)$ and
the four-graviton vertex by a factor $(1+\b_4)$, with constants
$\b_3$ and $\b_4$. For
$\b_3=\b_4=0$ we recover GR. Observe that, since $\b_3$ and $\b_4$ are
defined with respect to a given gauge choice, they are gauge-invariant
by definition. This is in fact the same logic used to define in a
gauge-invariant manner the PPN parameters $\bppn$ and $\gppn$.

We then use a Feynman diagram approach to compute the 
modifications induced by $\b_3$
and $\b_4$ on various
observables in classical GR. Diagrammatic approaches and 
field-theoretical methods have been in use in classical GR for 
a long time, see e.g.~\cite{DF96,Damour:2001bu,Blanchet:2003gy}. 
We make use of 
the effective field theory formulation proposed 
in~\cite{Goldberger:2004jt},  which provides a clean
and systematic separation of the effects that depend 
on (model-dependent) short-distance physics from long-wavelength
gravitational dynamics,
and in the
non-relativistic limit (after performing a multipole expansion)
has manifest power counting in the typical velocity
$v$ of the source.

In the next section we see how these deformed vertices give
additional terms in the PN effective Lagrangian. In particular we
find that, in the conservative sector of the theory, the
introduction of $\b_3$ is phenomenologically equivalent, at 1PN level, to the
introduction of a non-trivial value of $\bppn$ given by
$\bppn=1+\b_3$. However, $\b_3$ also affects the radiative sector of
the theory, i.e. the Lagrangian describing the interaction between the
matter fields and the gravitons radiated at infinity.

Before entering into the technical aspects, however, let us clarify
the  meaning of the introduction of $\b_3$ and $\b_4$. 
In ordinary GR, with $\beta_{3,4}=0$, coordinate transformation
invariance ensures that the negative norm states decouple. After gauge fixing
(in the De~Donder gauge for instance), the kinetic terms for all of the ten 
components of the metric are invertible, but four of them have the 
wrong sign, i.e. they give rise to negative norm states. 
In the De Donder gauge
the six positive-norm states that
diagonalize the kinetic term  are 
\be
\tilde{h}_{ij}\equiv h_{ij}+\frac
12\delta_{ij}(h_{00}-\delta^{lm}h_{lm})\, ,
\ee
while the four ``wrong-sign'' components are given by the spatial
vector $h_{0i}$ and by the scalar
\be
\tilde{h}_{N}\equiv h_{00}-\delta^{lm}h_{lm}\, .
\ee
In standard GR the existence of these negative-norm states do not create
difficulties because they are coupled to 
four integrals of motion (energy, 
and the three components of angular momentum), so they cannot be
produced. In contrast,
the remaining six ``healthy''
components couple to the source multipole moments. 
After complete gauge fixing one finds that among the six positive-norm 
states, four obey Poisson-like equations, so they do not radiate
(even though they are non-radiative physical 
degrees of freedom), while the remaining two 
are the radiative degrees of freedom representing 
GW's \cite{Flanagan:2005yc}.

Allowing $\beta_3\neq 0$ has the effect that the negative
norm state $\bar h_{N}$ now
couples, already at lowest order, to a non conserved quantity, 
namely to a combination of the Newtonian kinetic and potential energy 
of the binary system. 
This means that, in general, a modification of GR in which we just
change the strength of the three-graviton vertex cannot be taken as a
fundamental field theory, neither at the quantum level, nor even at
the classical level, since the negative-norm state  contributes to
the classical radiated power (a related concern is that, for
$\beta_3\neq 0$, the energy--momentum tensor is in general not conserved).
A consistent classical and quantum field
theory could in principle emerge from a simultaneous modification of
all the vertices of the theory, such as the three-, four- and
higher-order graviton vertices, together with a related modification of the
graviton-matter couplings. As a trivial example, an overall rescaling
of the gauge coupling in a Yang-Mills theory, or of Newton's constant
in GR,  results in a combined
modification of all the vertices, but obviously introduces no
pathology. Anyway, our approach to the problem is purely
phenomenological. We introduce $\b_3$ 
and $\b_4$ simply as ``tags'' that allow us to track the
contribution of the three- and four-graviton vertices throughout the
computations. As long as $|\b_3|\ll 1$ and $|\b_4|\ll 1$, the
corrections that they induce to the radiated power are small compared
to the standard GR result, so the total radiated power is given by the
GR result plus a small correction, and in particular the total
radiation emitted  is positive. 
At this phenomenological level the introduction of modified vertices
is therefore acceptable, and provides a simple and, most importantly, gauge
invariant manner of quantifying how well different observations
constrain the non-linear sector of GR, in a way which is intrinsic to
GR itself, without reference to any other specific field theory.

In this sense, our approach is  close in spirit to the
phenomenological PPN approach, and can be seen as an extension of it
where the radiative sector of the theory is  also modified.
Another approach which is related to ours is the one proposed
in ref.~\cite{Arun:2006yw,Arun:2006hn}. They consider
the phase of the GW emitted during
the coalescence of compact binaries, which up to 3.5PN has the
form
\be\label{Psipsik}
\Psi(f)=2\pi f t_c-\Phi_c+
\sum_{k=0}^7[\psi_k+\psi_{kl}\ln f] f^{(k-5)/3}\, ,
\ee
where $f$ is the GW frequency, and $t_c$ and $\Phi_c$ are the time and the
phase at merger.
The seven non-zero coefficients $\psi_k$ with $k=0,2,3,\ldots 7$ and the
two non-zero
coefficients $\psi_{kl}$ with $k=5,6$ are known from the PN
expansion, in terms of the two masses $m_1$ and $m_2$. 
In  ref.~\cite{Arun:2006yw,Arun:2006hn} 
they study how the template is affected
if these coefficients are allowed to vary,
so that two of them, the 0PN coefficient $\psi_0$ and
the 1PN coefficient $\psi_2$,
are used to fix the masses $m_1$ and $m_2$ of the two stars, while varying
any of the remaining coefficients with $k\geq 3$ provides a test of GR.
In the case of coalescing binaries at
interferometers, our introduction  of $\b_3$  and $\b_4$
is a particular case of a more general
analysis in which one treats  the quantities $\psi_k$ and $\psi_{kl}$
as free parameters, but it has a sharper field-theoretical meaning
since $\b_3$ and $\b_4$ measure 
the deviation of the three- and four-graviton vertices from
the GR prediction. For the same reason, we are also able to compare
the effect of $\b_3$ on the waveform of coalescing binaries  with its effect
on binary pulsar timing and on solar system experiments, while in the
phenomenological approach in which the parameters  $\psi_k$ and
$\psi_{kl}$ of the GW phase are taken as free parameters, a
modification of the waveform of coalescing binaries cannot be related
to a modification of the binary pulsar timing formula.

Another  issue is whether a modification of the vertices 
of this form (typically with $\b_3$, $\b_4$, etc. not independent, but
related to each other by some
consistency conditions) could
emerge from a plausible and consistent extension of GR. Actually,
a typical UV completion of GR
at an energy scale $\Lambda$ 
will rather generate corrections to the vertices that are
suppressed by inverse powers of $\Lambda$, so it would give rise to an energy
dependent $\beta_3$, e.g. $\beta_3=E^2/\Lambda^2$, which furthermore, at the
energy scales that we are considering and for any sensible choice of
$\Lambda$, would be utterly negligible. Still, let us remark that
this kind of behavior is not
a theorem. It assumes  the
UV-IR decoupling typical of effective field theories, and one can
exhibit counterexamples. For instance,
in non-commutative Yang-Mills theories there is a UV-IR mixing, such that 
low-energy processes receive contributions from loops where
very massive particles are running, and these contributions are independent 
of the mass of these particles~\cite{Minwalla:1999px}. Anyway, again  our
aim here 
is not to test any given consistent extension of GR, but rather provide 
a simple and phenomenologically
consistent way of quantifying how well various experiments can test
the non-linearities of GR, and  quantify  how the results of 
different experiments compare among themselves. 

It is also 
interesting to observe that, even when  $\b_3$ and $\b_4$
are non-zero, the graviton remains massless at the classical level, since 
 $\b_3$ and $\b_4$ affect interaction terms, but not the kinetic term. 
 The breaking of diffeomorphism invariance induced by  $\b_3$ 
and $\b_4$ could in principle generate a graviton mass at the one-loop level. 
However, even if we are using the language of quantum field theory, in the end we are only interested in the classical theory, since quantum loops are suppressed by powers of $\hbar/L$, where $L$ is the angular momentum of the system, so they are completely negligible for a macroscopic system.\footnote{At the quantum level, if a mass is generated, it
will be power divergent with the cutoff, and will have to  be fine tuned
order by order in perturbation theory.
However, the fact that quantum divergences
have to be subtracted order by order is a generic problem of the
standard quantum extension of GR, independently of  $\beta_3$. In any case, again, our approach is purely phenomenological, and the introduction of $\beta_3$ is simply a tool for tracking a specific contribution to the computation.}

\section{Effective  Lagrangian from a 
modified three-graviton vertex}\label{sect:comp}

To perform our computations we use the effective field theory 
formulation proposed in ref.~\cite{Goldberger:2004jt}.
Computations of the conservative dynamics at 2PN level have been performed 
using this effective field theory technique 
\cite{Chu:2008xm,Gilmore:2008gq}
and the results are in agreement with the classic 2PN results 
of refs.~\cite{DD81,DD82}, while the full 3PN result for non-spinning
particles to date has only been obtained with the standard
PN formalism using dimensional
regularization ~\cite{Damour:2001bu,Blanchet:2003gy}
(see ref.~\cite{Blanchet:2006zz}, or
Chapter~5 of  ref.~\cite{Maggiore:1900zz},
for a pedagogical introduction to the PN
expansion and for a more complete list of references).
Spin-spin contributions at 3PN order in the conservative
two-body dynamics  have recently been  
computed both with the effective field theory
techniques~\cite{Porto:2008tb,Levi:2008nh} and with the ADM 
Hamiltonian formalism~\cite{Steinhoff:2008ji} (see also
\cite{Galley:2008ih,Galley:2009px} for other applications of the EFT technique 
related to gravitational radiation).

\begin{figure}
\includegraphics[width=0.45\textwidth]{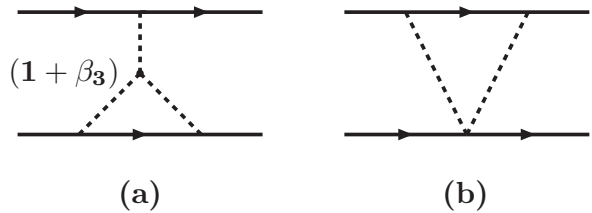}
\caption{\label{fig1}
The diagrams that give the terms proportional to $G_N^2$ in the 1PN
Lagrangian. Dashed lines denote potential gravitons, solid lines the
point-like sources.
}
\end{figure}

In the formalism of ref.~\cite{Goldberger:2004jt},
after integrating out length-scales
shorter than the size of the compact objects, the action 
becomes
\be
S=S_{\rm EH}+S_{\rm pp}\, , 
\ee
where $S_{\rm EH}$ is the
Einstein-Hilbert action and
\be
S_{\rm pp}=-\sum_{a}m_a\int d\tau_a\, 
\ee 
is the point particle action.
Here $a=1,2$ labels the two bodies in the binary system and
$d\tau_a=\sqrt{\gmn(x_a) dx_a^{\mu}dx_b^{\nu}}$.
One then observes that, in the binary problem, the gravitons appearing
in a Feynman diagram can be divided into two classes: the forces
between the two bodies with relative distance $r$ and relative speed
$v$ are mediated by
gravitons whose
momentum $k^{\mu}$ scales as $(k^0\sim v/r,|{\bf k}|\sim 1/r)$. These
are called ``potential gravitons'' and are off-shell, so they can
only appear in internal lines. The gravitons radiated to infinity
rather have
$(k^0\sim v/r,|{\bf k}|\sim v/r)$.
One then writes $\gmn=\emn+\hmn$ and 
separates $\hmn$ into two parts,
$\hmn(x)=\bhmn(x)+H_{\mu\nu}(x)$
with $\bhmn(x)$ describing the radiation gravitons and
$H_{\mu\nu}(x)$ the potential gravitons.  
One fixes the  de Donder gauge and,
expanding the action in powers of
$\bhmn$ and $H_{\bf k\mu\nu}(x_0)$, one can
read off the propagators and the vertices, 
and write down the Feynman rules of the theory. Then, using
standard methods from quantum field theory, one can
construct an effective Lagrangian that, used at tree level, reproduces
the amplitudes computed with the Feynman graphs. For a classical
system, whose angular momentum $L\gg\hbar$, only tree graphs
contribute, and reproduce the classical Lagrangian that is usually 
derived from GR using 
the PN expansion. For instance,
the conservative dynamics of the two-body problem, at 
1PN level, is given by the 
Einstein--Infeld--Hoffmann Lagrangian, 
\bees\label{2L2PN}
L_{\rm EIH}&=&\frac{1}{8}m_1 v_1^4+ \frac{1}{8}m_2 v_2^4
+\frac{G_Nm_1m_2}{2r}\, 
\[ 3 (v_1^2+v_2^2)\right.\nn\\
&&\hspace*{-4mm}\left. -7{\bf v}_1\bdot{\bf  v}_2
-(\hatr\bdot{\bf v}_1)(\hatr\bdot{\bf v}_2) -\frac{G_N(m_1+m_2)}{r}\].
\ees
In the language of the effective field theory of 
Ref.~\cite{Goldberger:2004jt}, this result is obtained from Feynman diagrams
involving the exchange of potential gravitons. In particular, the
terms linear in $G_N$ in \eq{2L2PN} are obtained from a single
exchange of potential gravitons between two matter lines, see Fig.~4 of 
Ref.~\cite{Goldberger:2004jt}, while the term proportional to $G_N^2$
is obtained from the sum of the two graphs shown in  Fig.~\ref{fig1}.
In the derivation of the conservative 1PN Lagrangian, 
the three-graviton vertex only enters through the graph 
in Fig.~\ref{fig1}a. Multiplying this vertex
by a factor $(1+\b_3)$ and repeating the same computation as in
ref.~\cite{Goldberger:2004jt} we get
the additional contribution to the conservative part of the Lagrangian
\begin{eqnarray}
\label{lb1}
\D{\cal L}_{\rm cons}=-\beta_3\frac{G_N^2 m_1 m_2 (m_1+m_2)}{r^2}\, .
\end{eqnarray}

\noindent

Comparing this result with  the Lagrangian  whose equations of motion
are the same as the equations of motion of a test particle in the PPN
metric (\ref{g00})--(\ref{gij}) (see eq.~(6.80) of ref.~\cite{Willbook})
we find that,  to 1PN order and
as far as the conservative dynamics is concerned, the introduction of $\b_3$
gives rise to
a PPN theory with $\b=1+\b_3$, i.e. $\bar{\b}=\b_3$, while  $\g=1$ as in GR. 
Therefore the bound on $\bar{\b}$ from the perihelion of
Mercury  translates 
into $|\b_3|<3\cdot 10^{-3}$, while
\eq{Cassini44} translates  into the bound
(at $68\%$ c.l.)
\be\label{lunar}
|\b_3|<2\cdot 10^{-4}\, .
\ee
This bound reflects the fact that $\b_3$, just as the PPN parameter $\bar{\beta}$, violates the weak equivalence principle.
The introduction of $\b_3$, however, also affects the radiative sector
of the theory, something that is not modeled in the phenomenological  PPN
framework since the latter by definition is only concerned with the
motion of test masses in a deformed metric, and therefore
only modifies the conservative part of the dynamics. 
Note however that, in the framework of multiscalar-tensor
theories, the extension of the PPN formalism introduced
in Ref~\cite{DF92}  allows for a consistent treatment of
both the conservative dynamics (including  the case of strongly
self-gravitating bodies) and of radiative effects.

\begin{figure}
\includegraphics[width=0.45\textwidth]{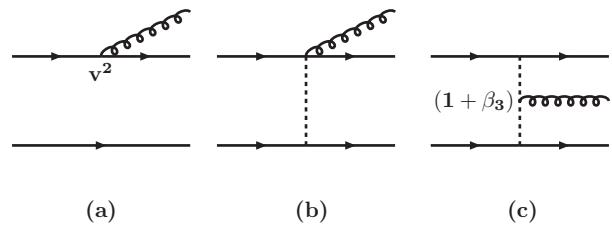}
\caption{\label{figGR6}
The diagrams that contribute to the matter-radiation Lagrangian.
Dashed lines denote potential gravitons, wiggly lines radiation
gravitons, and solid lines the point-like sources.}
\end{figure}

It is clearly interesting to see what bounds on $\b_3$ can
be obtained from experiments that probe the radiative sector of GR, such
as the timing of binary pulsars or the observation of the coalescence
of compact binaries at interferometers. 
The effective
Lagrangian describing the interaction 
of the binary system with radiation gravitons is obtained by computing
the three graphs in  Fig.~\ref{figGR6} (corresponding to 
Fig.~6 of ref.~\cite{Goldberger:2004jt}), and the introduction
of $\b_3$ affects the $HHh$ vertex in Fig.~\ref{figGR6}c. 

Computing
these graphs as in ref.~\cite{Goldberger:2004jt}, but with our modified 
three-graviton vertex, we find
\be\label{lquad}
{\cal L}_{\rm rad}=
\frac{1}{2\mpl}[Q_{ij}R_{0i0j}+qR_{0i0i}+ \beta_3 
(3 V h_{00} +Z^{ij} h_{ij})]\, ,
\ee
where $Q_{ij}$ is the quadrupole moment of the source and we define
\bees
q&=& \frac{1}{3} \sum_a m_a x_a^2\, , \\
V(r)&=& \frac{G_N m_1m_2}{r}\, ,\\
Z^{ij}(r)&=&\frac{G_Nm_1m_2r^ir^j}{r^3}\, ,
\ees
where ${\bf r}={\bf x}_1-{\bf x}_2$.
The  term $Q_{ij}R_{0i0j}$
in \eq{lquad} is the usual quadrupole interaction. 
The second term, $qR_{0i0i}$, is
non-radiating when $\b_3=0$, but we will see that
for $\b_3\neq 0$ it contributes to the radiated power when the orbit
is non-circular. The last two terms in \eq{lquad}
are the explicit
$\beta_3$-dependent terms induced by the modification of the
three-graviton vertex in Fig.~\ref{figGR6}c. 

Finally, we have omitted
a term where $h_{00}$ is coupled to the conserved energy (at this order) and 
$h_{0i}$ is coupled to the conserved angular momentum, since these
terms do not  generate gravitational radiation.

\begin{figure}
\includegraphics[width=0.3\textwidth]{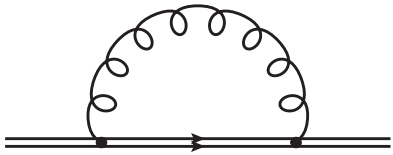}
\caption{\label{figGR3}
The self-energy 
diagrams whose imaginary part gives the radiated power. The wiggly
line can refer either to $h_{ij}$ or to $h_{00}$, and the vertices to
any of the four $I^a_{ij}S^a_{ij}$ with $a=1,\ldots,4$,
see the text.
}
\end{figure}

The back-reaction of GW emission on the source can be computed as usual
from the energy balance equation  
\be\label{balance}
P_{\rm GW}=-\dot{E}\, , 
\ee
where
$P_{\rm GW}$ is the power radiated in GWs and $E$ the orbital energy of
the system. To obtain 
the expression for the radiated power 
for $\b_3\neq 0$ we cannot simply use the quadrupole
formula of GR, since the introduction of 
$\b_3$ generates new contributions. To take them into
account we proceed as in ref.~\cite{Goldberger:2004jt}, and we compute
the imaginary part of the graph shown in 
Fig.~\ref{figGR3}. The vertices of the graph can be read from
\eq{lquad}. When $\beta_3=0$ the only relevant vertex comes from
$Q_{ij}R_{0i0j}$, and the computation of the imaginary part gives back
the Einstein quadrupole formula \cite{Goldberger:2004jt}. In our case we
have various possible vertices, and we must compute the imaginary part of
\be\label{imagine}
\frac{-i}{8\mpl^2}
\sum_{a,b=1}^4 \int dt_1dt_2\, I^a_{ij}(t_1)I^b_{kl}(t_2)\nn\\
\langle S^a_{ij}(t_1)S^b_{kl}(t_2)\rangle\, ,
\ee
where $I^a_{ij}=(Q_{ij},q\d_{ij},\beta_3 V\d_{ij},\beta_3  Z_{ij})$ 
depends on the matter variables
and
$S^a_{ij}=(R_{0i0j},R_{0i0j}, \d_{ij}h_{00},h_{ij})$ on the
gravitational field.
When both vertices of the diagram in 
Fig.~\ref{figGR3} are proportional to the quadrupole, one obtains the
usual GR result 
\be
\label{emqq}
P_{QQ}=\frac{G_N}{5}\langle \dddot{Q}_{ij}\dddot{Q}_{ij}\rangle\, ,
\ee
as already found in~\cite{Goldberger:2004jt}. Computing the other
contributions we find that
the terms $P_{Qq}$ and $P_{qq}$ vanish identically. In fact,
the  $Qq$ and $qQ$ graphs 
vanish because $Q_{ij}$ is traceless, while the $qq$ graph vanishes
because $\delta_{ij}\delta_{kl}$ gives zero when contracted
$\delta_{ik}\delta_{jl}+\delta_{il}\delta_{jk}-\frac 23\delta_{ij}\delta_{kl}$,
which is the tensor that comes out from the two-point function 
$\langle R_{0i0j}R_{0k0l}\rangle$.
The $QV$, $qZ$ and $VZ$ graphs vanish for similar reasons, so
the only relevant contributions come
from the $QZ$ and $qV$ graphs, and we find
\be
\label{exem1}
P_{QZ}=-2\beta_3G_N\langle\dddot Q_{ij}\dot Z_{ij}\rangle\, ,
\ee 
and
\be
\label{exem2}
P_{qV}=-6\beta_3G_N\langle\dddot q\dot V\rangle\, . 
\ee
As for the $VV$ and $ZZ$ graphs, they give a
contribution that, from the point of view of the multipole expansion,
is of the same order as the quadrupole radiation but  proportional
to $\b_3^2$, and can be neglected.

We can now use these results to perform the comparison with binary
pulsars and with interferometers.

\section{Comparison with experiment}\label{sect:exp}

As we already saw in \eq{lunar}, solar system experiments, and 
in particular  lunar laser ranging, give the bound
$|\b_3|<2\cdot 10^{-4}$. In this section we study the bounds on $\b_3$
that can be obtained from binary pulsars and from the detection of
coalescing binaries at interferometers.  

\subsection{Binary pulsars}

Since  $\beta_3$ modifies the emitted power already at Newtonian
order, the energy of the orbit in
\eq{balance} can now be directly
computed using the Keplerian equations of motion. 
We see that this test of GR is conceptually different from the tests based
on solar system experiments. The latter only probe the conservative
part of the Lagrangian, i.e. the $\b_3$-dependent term given in
\eq{lb1}, 
while binary pulsars are sensitive to the
$\b_3$ dependence given in the radiation Lagrangian (\ref{lquad})
(even if the effect of $\beta_3$ on the conservative dynamics will
also enter, 
through the determination of the masses of the stars from the
periastron shift, see below).

Using the Keplerian equations of motion for an elliptic orbit of
eccentricity $e$ we get
\begin{eqnarray}
\label{pellips}
P_{QQ}&=&\frac {32 G_N^4 \mu^2 M^3}{5 a^5(1-e^2)^{7/2}}
\left(1+\frac{73}{24} e^2 +\frac{37}{96}e^4\right)\,,\\
P_{QZ}&=&\beta_3 \frac {32 G_N^4 \mu^2 M^3}{5 a^5(1-e^2)^{7/2}}
\left(\frac{5}{2}+\frac{175}{24}e^2+\frac{85}{96}e^4\right),\\
P_{qV}&=&-\beta_3 \frac {32 G_N^4 \mu^2 M^3}{5 a^5(1-e^2)^{7/2}} 
\left(\frac{5}{16}e^2+\frac{5}{64}e^4\right)\, ,
\end{eqnarray}
where $M=m_1+m_2$ is the total mass, $\mu=m_1m_2/M$ is the reduced mass,
and we will also use the notation
$\nu =m_1m_2/M^2$ for the symmetric mass ratio.
From the energy balance equation we then get the evolution of the
orbital period $P_b$,
\begin{eqnarray}
\label{dotT}
\frac{\dot{P}_b}{P_b}=-\frac{96}{5}G_N^{5/3}\nu\, M^{5/3}
\left(\frac{P_b}{2\pi}\right)^{-8/3}[f(e)+\beta_3 g(e)]\,,
\end{eqnarray}

\noindent
where
\begin{eqnarray}
\label{fege}
&&f(e)=\frac{1}{(1-e^2)^{7/2}}
\left(1+\frac{73}{24} e^2 +\frac{37}{96}e^4\right)\,,\\
&&g(e)=\frac{1}{(1-e^2)^{7/2}}\left(\frac{5}{2}+\frac{335}{48}e^2+
\frac{155}{192}e^4\right)\,.
\end{eqnarray}
The term proportional to $f(e)$ is the standard GR
result~\cite{Peters}, while the term proportional to $g(e)$ is the
extra contribution due to $\b_3$.

In order to measure $\beta_3$ from the dynamics of
binary pulsars, we must also determine the  
dependence on $\b_3$ of the  periastron shift $\omega$ and of the 
Einstein time delay $\gamma_E$, since these two observables are used to
determine the masses of the two compact stars. 
In particular the periastron shift 
fixes the total mass $M$ of the system, while the Einstein
time delay $\g_E$ measures a different combination of masses,
see e.g. eqs.~(6.56) and (6.93) of ref.~\cite{Maggiore:1900zz}.
Using the conservative Lagrangian with the modification (\ref{lb1})
and repeating the standard textbook computation of the 
periastron shift $\omega$,
we find that the value $\omega_{\b_3}$
computed in a theory with $\b_3\neq 0$ is related to the GR value 
$\omega_{\rm GR}$ by
\be
\omega_{\beta_3}=\(1-\frac{\beta_3}{3}\)\omega_{\rm GR}\, , 
\ee
while
the Einstein time delay is unchanged because it is
not affected by the post-Keplerian parameters. So, if $\b_3\neq 0$, the true
value of the total mass $M$ of the binary system, that enters in 
\eq{dotT}, is not the one that would be inferred from the  periastron
shift using the predictions of GR, but rather we get
\be
M_{\beta_3}=\(1+\frac{\beta_3}{2}\)M_{\rm GR}\, . 
\ee
Similarly,  
using  eq.~(6.93) of ref.~\cite{Maggiore:1900zz},
for the symmetric mass ratio $\nu$ we get 
\be
\nu_{\beta_3}=(1+w\beta_3)\nu_{\rm GR}\, ,
\ee 
where 
\be
\label{nubeta}
w=\frac{\kappa}{3}\frac{\sqrt{1+4\kappa}-2}{\sqrt{1+4\kappa}}
\frac 1{(1+4\kappa)^{1/2}-(1+\kappa)}
\ee
and 
\be
\kappa=\frac{\gamma}{e}\, \(\frac{2\pi}{P_b}\)^{1/3}\,
(G_NM_{\rm GR})^{-2/3}\, . 
\ee
Putting everything together and keeping only the
linear order in $\beta_3$ we finally find that the ratio between the
value of $\dot{P}_b$ computed at $\b_3\neq 0$ and the value of
$\dot{P}_b$ computed in GR is
\be
\frac{\dot{P}_b^{(\beta_3)}}{\dot{P}_b^{\rm GR}}=1+\beta_3 \tilde{g}(e)\, ,
\ee
where 
\be
\tilde{g}(e)=\frac{g(e)}{f(e)}+ \frac{5}{6}+w\, .
\ee
Observe that the term ${g(e)}/{f(e)}$ comes from the  effect of $\b_3$
on the radiative sector of the theory, while the term $(5/6)+w$ comes
from the effect of $\b_3$ on the conservative sector, i.e. on the mass
determination. 
Inserting the numerical values for the Hulse-Taylor pulsar, we
get $\tilde{g}(e)\simeq 3.21$. Note that
${g(e)}/{f(e)}\simeq 2.38$, so $\tilde{g}(e)$
is  dominated by the  effect of $\b_3$
on the radiative sector of the theory.
For this binary pulsar, after correcting for
the Doppler shift due to the relative velocity between us and the
pulsar induced by the differential rotation of the Galaxy, the ratio
between the observed value   $\dot{P}_b^{\rm obs}$ and the GR prediction
$\dot{P}_b^{\rm GR}$ is 
$\dot{P}_b^{\rm obs}/\dot{P}_b^{\rm GR}=1.0013(21)$. Interpreting this as
a measurement of $\b_3$ we finally get 
$3.21\b_3=0.0013(21)$, i.e.
\be\label{boundb3}
\b_3=(4.0\pm 6.4)\cdot 10^{-4}\, ,
\ee
so the three-graviton vertex  
is consistent with the GR prediction  at the
0.1\% level. This bound is slightly worse, but
comparable, to the one from lunar laser ranging, \eq{lunar}. It should be
stressed, however, that \eq{boundb3} is really a test involving the
radiative sector of GR, while \eq{lunar} only tests the conservative sector.
For comparison, observe that in the
Standard Model the triple gauge boson couplings are measured to an
accuracy of about 3\%~\cite{Schael:2004tq}.

For the double pulsar we find
$\tilde{g}(e)\simeq 3.3$. Since $\dot{P}_b$ for the double pulsar 
is presently measured at the $1.4\%$ level \cite{Kramer:2006nb}, 
we get a larger bound compared to \eq{boundb3}. However, further monitoring of 
this system is expected to bring the error on $\dot{P}_b$  down to the 
$0.1\%$ level.

\subsection{Binary coalescences at interferometers}

We now compare these results with
what can be expected from the detection of a binary coalescence at GW
interferometers. 
In this case one can determine the physical parameters of 
the inspiraling bodies, by performing matched filtering of theoretical 
waveform templates.
In the matched filtering method any difference in the time behavior between
the actual signal and the theoretical template model will eventually
cause the two to go out of phase, with a consequent drop in the 
signal-to-noise ratio (SNR). The introduction of $\b_3$ and $\b_4$
affects the template, and in particular the accumulated phase 
\be
\phi=2\pi \int_{t_{\rm min}}^{t_{\rm max}} \!\!\!\!f(t) dt\,,
\ee
where $f(t)$ is the time-varying frequency of the source, and
the subscript min (max) denotes the values when the signal enters (leaves)
the detector band-width.
Thus in principle a detection of a GW signal from coalescing binaries
could be translated into a
measurement of the three- and four-graviton vertices. In this section
we investigate the accuracy of such a determination.

With respect to the timing of binary pulsars, there are at least three
important qualitative differences that affect the  accuracy at which
these systems can test the non-linearities of GR. First,
coalescing compact binaries in the last stage of the coalescence reach
values of $v/c\sim 1/3$, and are therefore much more relativistic than
binary pulsars, which rather  have $v/c\sim 10^{-3}$. Second,
the leading Newtonian result  for $\phi$
is  of order $(v/c)^{-5}$ so it is much larger than one, and to get the phase
with a precision $\D\phi\ll 1$, as needed by interferometers, all
the corrections at least up to $O(v^6/c^6)$ to the Newtonian result
must be included, so higher-order 
corrections are important even if they are numerically small
relative to the leading term.  In other words, even if PN corrections are
suppressed by powers of $v/c$ with respect to the leading term, they
can be probed up to high order
because what matters for GW interferometers is the overall
value of the PN corrections to the phase, 
and not their value relative to the large
Newtonian term. 
These two considerations should suggest that
interferometers are much more sensitive than pulsar timing to the
non-linearities of GR. 

On the other hand, for binary pulsars we can
measure not only the decay of the orbital period due to GW emission,
but also several other Keplerian observables,
that provide a determination of the geometry of the orbit, as well as
post-Keplerian observables, 
such as the periastron shift and the Einstein
time delay, which fix
the masses of
the stars in the binary system. This is not the case for the detection
of coalescences at interferometers. With interferometers the
parameters that determine the waveform, such as the
masses and spins of the stars, must be
determined from the phase of the GW itself, and one must then 
carefully investigate the
degeneracies between the determination of $\b_3$ (or of $\b_4$) and
the determination of the masses and spins of the stars. 
This effect clearly goes in the
direction of degrading the accuracy of parameter reconstruction
at GW interferometers, with
respect to binary pulsar timing, so in the end it is not obvious a
priori which  of the two,  GW interferometers or binary pulsar timing,
is more sensitive
to the non-linearities of GR. This question
is answered in what follows.

Repeating with $\b_3\neq 0$
the standard computation of the orbital
phase for a circularized orbit,
we find that  $\b_3$  
modifies the orbital phase $\phi(t)$ 
already at 0PN (i.e. Newtonian) level, where to linear order in
$\b_3$ we get
\be\label{phi0PN}
\phi^{\rm 0PN}=-\frac{\Theta^{5/8}}{\nu} (1+b_0\beta_3)\, ,
\ee
with $b_0=-5/2$, and $\Theta$ is defined as
\be\label{defTheta}
\Theta =\frac{\nu (t_c-t)}{5GM}\, (1-b_0\beta_3)\, ,
\ee 
where $t_c$ is the time of coalescence.
Combining the factors $\nu$ and $M$ which enter in the definition of $\Theta$
with the explicit factor $1/\nu$ in \eq{phi0PN} we recover the well-know result
that the Newtonian phase depends on the masses of the stars only
through the chirp mass $M_c=\nu^{3/5}M$. From \eq{phi0PN} we
immediately understand the crucial role that degeneracies have for
interferometers. In fact, since $M_c$ is determined from \eq{phi0PN}
itself, using only the 0PN phase (\ref{phi0PN}) it is impossible to
detect the deviation from the prediction of GR induced by $\b_3$. A
non-zero value of $\b_3$ would simply induce an error in the
determination of $M_c$. 

The same happens at 1PN level. In fact, at 1PN order and with
$\b_3\neq 0$ 
the phase has the general form
\be
\label{1PN}
\phi^{\rm 1PN}=-\frac{\Theta^{5/8}}{\nu}\[ (1+b_0\beta_3)
+a_1(\nu)(1+b_1\b_3 ) \Theta^{-1/4}\]
\, ,
\ee
where, as before, 
$b_0=-5/2$ is the 0PN correction proportional to $\b_3$, while
\be\label{a1nu}
a_1(\nu)=\frac{3715}{8064} + \frac{55}{96}\nu
\ee 
is the 1PN GR prediction \cite{Blanchet:2006zz,Maggiore:1900zz},
and 
$b_1$  (which is possibly $\nu$-dependent) parametrizes the 1PN correction
due to $\b_3$. (For
simplicity, we only wrote explicitly 
the term linear in $\b_3$ since $|\b_3|$ is 
much smaller than one, but all our considerations below can be
trivially generalized to terms quadratic in $\b_3$, just by allowing the
function $b_1(\nu)$ to depend also on $\b_3$). In general we expect
$b_1$ to be $O(1)$, and we will see below that for our purposes this
estimate is sufficient.

Using $M_c$ and $\nu$ as independent mass
variables, in place of $m_1$ and $m_2$, we see that while
the effect of $\b_3$ on the 0PN phase can be reabsorbed into $M_c$,
its effect on the 1PN phase can be reabsorbed into a rescaling of
$\nu$. Observe that, in the detection of a single coalescence event,
GW interferometers do not measure the functional dependence of $\nu$ of
the 1PN phase, but only its numerical value for the actual value of
$\nu$ of that binary system, so we cannot infer  the presence of a term
proportional to $\b_3$ from the fact that it changes the functional
form of the $\nu$-dependence from the one obtained by \eq{a1nu}.
Thus, even at 1PN order, 
it is impossible to detect the deviations from GR
induced by $\b_3$. A non-zero $\b_3$ would simply induce an error
on the determination of $M_c$ and $\nu$, i.e. on the masses of the two
stars. 

We then examine the situation at 1.5PN
order. Let us at first neglect the 
spin of the two stars. Then
the 1.5PN phase with $\b_3\neq 0$ has the generic form
\bees
\label{1.5PN}
\phi^{\rm 1.5PN}&=&-\frac{\Theta^{5/8}}{\nu}\[ (1+b_0\beta_3)
+a_1(\nu)(1+b_1\b_3 ) \Theta^{-1/4}\right.\nn\\
&&\left.\phantom{-\frac{\Theta^{5/8}}{\nu}} 
+a_2(1+b_2\b_3) \Theta^{-3/8}
\]
\, ,
\ees
where $a_2=-3\pi/4$ is the 1.5PN GR prediction and $b_2$ is the
(possibly $\nu$-dependent) 1.5PN correction
due to $\b_3$. Again, we will not need its exact value, and we will
simply make the natural assumption that it is $O(1)$.

However, for the purpose of determining $\b_3$,
neglecting  spin is not correct. Indeed, 
at 1.5PN order  the spin of the bodies enters 
through the spin-orbit coupling, and
the evolution of the GW frequency $f$ with time is given 
by~\cite{Kidder:1992fr}
\be\label{dfdt15PN}
\frac{df}{dt}=\frac{96}{5}\pi^{8/3} M_c^{5/3} f^{11/3}
\[ 1-\frac{24}{5} a_1(\nu) x +(4\pi -\b_{\rm LS})x^{3/2}\],
\ee
where $x=(\pi Mf)^{2/3}$, while $\b_{\rm LS}$ describes the spin-orbit
coupling  and is given by
\be
\b_{\rm LS} =\frac{1}{12}\sum_{a=1}^2\[
113\frac{m_a^2}{M^2}+75\nu\]{\bf \hat{L}}\bdot\vchi_a\, ,
\ee
where ${\bf L}$ is the orbital angular momentum, 
$\vchi_a={\bf S}_a/m_a^2$ and
${\bf S}_a$ is the  spin of the $a$-th body.
In principle $\b_{\rm LS}$ evolves
with time because of the precession of ${\bf L}$, ${\bf S}_1$ and
${\bf S}_2$. However, it turns out that in practice it is almost
conserved, and can be treated as a
constant~\cite{Cutler:1994ys}. Integration of \eq{dfdt15PN} then
shows that, in the 1.5PN phase, $\b_{\rm LS}$ is exactly degenerate
with $\b_3$ in \eq{1.5PN}. Furthermore, observe that 
$\b_{\rm LS}$, depending on the spin configuration, can reach a
maximum value of about 8.5~\cite{Cutler:1994ys} (and its maximum value
remains large
even in the limit $\nu\ra 0$), while $\b_3$ is
already bound by laser ranging at the level of $2\times 10^{-4}$ and by
pulsar timing at the level of $10^{-3}$ 
(which tests the radiative sector, as do GW interferometers). Thus, 
the effect of $\b_3$ at 1.5PN is simply reabsorbed into a (very small)
shift of $\b_{\rm LS}$.

At 2PN order $\b_3$ is
degenerate with the parameter $\s$ that describes the spin-spin
interaction
\bees
\s &=&\frac{\nu}{48}
[721(\hat{\bf L}\bdot\vchi_1)-247\vchi_1\bdot\vchi_2
(\hat{\bf L}\bdot\vchi_2) ]\nn\\
&&+\frac{1}{96}\sum_{a=1}^2\frac{m_a^2}{M^2}
\[719 (\hat{\bf L}\bdot\vchi_a)^2-233\vchi_a^2\]\, .
\ees
The first term is the one which is usually quoted in the literature,
first computed in \cite{Kidder:1992fr} (see also
\cite{Poisson:1995ef, Blanchet:1995ez}). The term in the second line,
computed recently in \cite{Racine:2008kj},
is however of the same order, and must be included.

The first term is proportional to $\nu$, and reaches a maximum value
$\s_{\rm max}(\nu)\simeq 10\nu$. In a coalescence with
very small value of $\nu$, this term is therefore suppressed; e.g. in  
an extreme mass-ratio inspiral (EMRI) event at LISA
where a BH of mass $m_1=10\msun$ falls into a supermassive BH with
$m_2=10^6\msun$, one has $\nu=10^{-5}$ and   the term in the first line
has a maximum value $\sim 10^{-4}$.
If this standard term gave  the full answer, 
a value of $\b_3$ in excess of this value could therefore
give an effect that cannot be ascribed to $\s$. However, the
presence of the new term recently computed in 
\cite{Racine:2008kj} spoils this reasoning, since it is not
proportional to $\nu$. The conclusion is  that, just as with
$\b_{\rm SL}$ at 1.5PN order, the effect of $\b_3$ at 2PN order is
just reabsorbed into a small redefiniton of $\s$, and therefore simply
induces an error in the reconstruction of the spin configuration
(observe also that fixing $\b_{\rm LS}$ does not allow us to fix the
spin combinations that appear in $\s$.) 

One could in principle investigate the effect of $\b_3$ on 
higher-order coefficients, such as the 2.5PN term $\psi_5$ and the 3PN
term $\psi_6$ in \eq{Psipsik}, which at LISA
can be measured with a precision of order 
$10^{-2}$~\cite{Arun:2006hn}. Unfortunately, it is difficult to
translate them into bounds on $\b_3$ because
at 2.5PN order one  finds that $\b_3$ is degenerate with a
different combination of spin and orbital variables, which is not
fixed by the 1.5PN spin-orbit term (see  Table~II of
ref.~\cite{Blanchet:2006gy}), while  
at 3PN order  the spin contribution is not yet known.
The conclusion is therefore that
interferometers cannot measure the three- and higher-order graviton
vertex, since the effect of a modified vertex is simply reabsorbed
into the determination of the masses and spin of the binary system.
The conclusion that interferometers 
are not competitive with pulsar timing for measuring deviations from GR
was also 
reached in Ref.~\cite{DF98}, although in a different   context. 
In fact, Ref.~\cite{DF98} was  concerned
with multiscalar-tensor theories, whose leading-order effect
is the  introduction of a term corresponding to dipole radiation (a
``minus one''-PN term).

\begin{figure}
\includegraphics[width=0.4\textwidth]{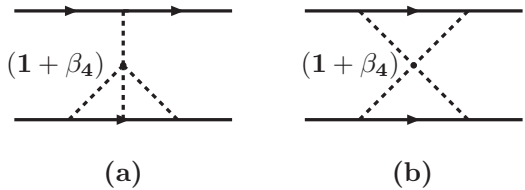}
\caption{\label{fig4}
The diagrams that contribute to the conservative dynamics, which are
affected by a modification of the four-graviton vertex.}
\end{figure}

\begin{figure}
\includegraphics[width=0.2\textwidth]{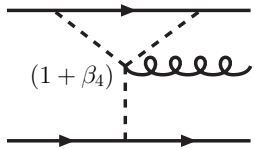}
\caption{\label{fig5}
The diagram that contributes to the radiative dynamics, which is
affected by a modification of the four-graviton vertex.}
\end{figure}

We now examine what can be said about the four-graviton vertex,
parametrized by $\b_4$.
In the conservative part of the
Lagrangian $\b_4$ contributes through the diagrams of Fig.~\ref{fig4}.
However, these contributions  only affect the equations of motion at
2PN order, so there is no hope to see them in solar system
experiments, where the velocities at play are very small.
For the same reason, no significant bound can be obtained from 
binary pulsars; from a simple order of magnitude estimate we find
that the Hulse-Taylor pulsar can only give a limit  $\b_4<O(10)$,
which is not significant. 

At GW interferometers,  $\b_4$ enters into
the phase for the first time at 1PN order, through the diagram in
Fig.~\ref{fig5}. However, it suffers exactly of the same degeneracy
issues as $\b_3$, so it cannot be measured to any interesting accuracy
at present or future interferometers, at least
with the technique discussed here.
Note however that in this paper we have worked in the restricted PN
approximation, in which only the harmonic at twice the source
frequency is retained. Higher-order harmonics 
however  break degeneracies between various parameters in the template~\cite{VanDenBroeck:2006ar}, and it would be interesting
to investigate  whether their inclusion  in the analysis allows one to put a bound on
  $\b_3$ and $\b_4$ from binary coalescences.

\section{Conclusion}\label{sect:concl}

We have proposed to quantify the accuracy by which various experiments
probe the non-linearities of GR, by translating their results  
into measurements of the
non-abelian vertices of the theory, such as the three-graviton vertex
and the four-graviton vertex. This is similar in spirit to tests of
the Standard Model of particle physics, where the non-abelian vertices
involving three and four gauge bosons have been measured at LEP and at
the Tevatron.

We have shown that, at a phenomenological level, this can be done in a
consistent and gauge-invariant manner, by introducing
parameters $\b_3$ and $\b_4$ that quantify the deviations from the GR
prediction of the
three- and four-graviton vertices, respectively. 
We have found that, in the conservative sector of the
theory, i.e. as long as one neglects the emission of gravitational
radiation at infinity, the introduction of $\b_3$ at 1PN order is phenomenologically
equivalent to the
introduction of a parameter $\bppn=1+\b_3$ in the parametrized PN
formalism. Strong bounds on $\b_3$ therefore come from solar system
experiments, and most notably from lunar laser ranging, that provides
a measurement at the $0.02\%$ level.

The modification of the three-graviton vertex however also affects the
radiative sector of the theory, and we have found that the timing of
the Hulse-Taylor pulsar gives a bound on $\b_3$ at the 0.1\% level, 
not far from the one
obtained from lunar laser ranging. Conceptually, however, the two
bounds have different meanings, since lunar laser ranging 
only probes  the conservative sector of the theory, while pulsar timing
is also sensitive  to the radiative sector. 

We have then studied the
results that could be obtained from the detection of coalescences at
interferometers, and we have found that, even if $\b_3$ already
modifies the GW phase at the Newtonian level and $\b_4$ at 1PN order,
their effect can always be
reabsorbed into other parameters in
the template, such as the mass and spin of the two bodies so,
rather than detecting a deviation from the GR prediction, one would
simply make a small error in the estimation of these parameters.

\vspace{5mm}\noindent
{\em Acknowledgments}. The work of UC, SF, MM and HS 
is supported by the Fonds National Suisse.
We thank Alvaro de Rujula for a
stimulating discussion and Thibault Damour
and the referees for useful comments. HS would
like to thank Steven Carlip and Bei-Lok~Hu, and UC
would like to thank Yi-Zen~Chu, for helpful discussions.

\end{document}